# DagFlow-AI: Agentic Synthetic Data Generation for Personalized Statistics Learning

**David Arthur[1*]**, **Ranjini Grove[2]**, **Zaher Kmail[1]**, **Curtis Atkisson[3]**
[1]Sciences and Mathematics, University of Washington Tacoma
[2]Statistics, University of Washington Seattle
[3]eScience Institute, University of Washington

## Abstract

To provide personalized learning experiences in courses that teach data analysis skills, instructors often search for datasets that appeal to a student's interests, align with specific learning objectives, and are appropriate for the student's skill level. Datasets that match all three criteria can be difficult to find. DagFlow-AI is an agentic AI data generation tool that creates realistic, synthetic datasets tailored to these criteria. Rather than mimicking existing datasets or generating observations directly, DagFlow-AI uses the broad knowledge and reasoning capabilities of large language models (LLMs) to specify the components of transparent, reproducible data-generating mechanisms from which observations are generated programmatically. We illustrate how the tool can be used to construct a realistic, synthetic dataset and describe how it aligns with both the Guidelines for Assessment and Instruction of Statistics Education (GAISE) and the Problem, Plan, Data, Analysis, and Conclusion (PPDAC) framework for teaching data analysis.

*Corresponding author. Email: dbarthur@uw.edu

# 1. INTRODUCTION

As data becomes an increasingly integral part of nearly every profession, higher education must equip students with the skills needed to analyze and reason about data. These learning experiences can be especially beneficial when students work with datasets that reflect their personal or career interests while also aligning with the learning objectives for the course or an assignment (Medina and Dogucu, 2026). However, finding a real dataset that is both relevant to students and well-suited to a particular learning objective can be difficult. Available datasets may lack the necessary variables or statistical properties, while those that appeal to one group of students may have little connection to the interests or fields of study of others. Widely used datasets may also have solutions, code, and analyses readily available online, limiting their suitability for projects and assessments.

Synthetic data provide an alternative. In this paper, we use the term *synthetic data* to refer to simulated observations that are designed to resemble data that could plausibly arise in the real world but are not collected from actual individuals, objects, or events. Because the context, variables, distributions, and relationships can be specified in advance, synthetic datasets can be designed to exhibit the statistical properties needed for a particular learning objective. The same statistical task can also be embedded in different subject areas, creating opportunities to tailor assignments to students' interests. However, producing a high-quality synthetic dataset requires translating a chosen topic and learning objective into a coherent set of variables and relationships, selecting appropriate distributions and functional forms, and verifying that the resulting data behave as intended. When performed manually, this process can require substantial time and statistical expertise.

Deep learning methods and large language models (LLMs) offer possible ways to facilitate this process. These tools can learn complex patterns and relationships from existing datasets and generate new observations that reproduce them, to varying degrees of success. Although this approach can produce realistic synthetic data, it does not fully address the other goals described above. Most methods require instructors to first identify a suitable real dataset on which to train the model, and they provide limited ability to tailor the resulting data to specific learning objectives.

An alternative use of LLMs is to leverage their broad knowledge and reasoning abilities to support the construction of realistic synthetic datasets. Rather than using LLMs to generate observations directly, we propose using an agentic AI framework to replicate and automate the conceptual work a person might perform when carefully constructing a dataset by hand. The process is divided into a series of tasks – including identifying variables, specifying their relationships in a directed acyclic graph, selecting distributions and defining model formulas – each managed by a specialized agent. The agents produce the structured components of an explicit data-generating model, while validation and data simulation are handled programmatically.

Although the basic idea is straightforward, implementing a reliable agentic workflow requires careful design. Prompts must be carefully constructed to ensure that each agent has clear instructions as well as the necessary context required for them to perform their task reliably. While performance of an agent can depend on the model being used, our experience has shown that changing models does not degrade performance, in particular when switching from older to newer models. However, even the best models using carefully constructed prompts can still produce inaccurate or unsupported output or may not always follow instructions consistently. Safeguards and validation procedures are therefore needed, including decisions about which tasks should be handled deterministically through code and which benefit from the flexibility of an LLM. The agents must also be able to exchange information and coordinate their work so that their individual outputs contribute to a coherent data-generating model.

In this work, we present DagFlow-AI, a tool that uses agentic AI to implement a synthetic-data generation workflow. The workflow is organized into six stages: (1) information gathering, (2) variable creation, (3) directed acyclic graph creation, (4) distribution selection, (5) formula generation, and finally (6) data simulation. At each of these stages, users can inspect the output and provide feedback to help ensure that the resulting dataset meets their needs. Through this iterative process, DagFlow-AI can create personalized, realistic datasets aligned with specific learning objectives, while reducing the time and technical effort required of users.

Section 2 of the paper provides the background on synthetic data generation for pedagogical purposes and motivates the development of DagFlow-AI. Section 3 describes the tool and its synthetic-data generation workflow. Section 4 presents a case study to demonstrate the creation of a personalized dataset for several common introductory statistics learning objectives. Finally, Section 5 provides a brief discussion of the tool, its limitations and directions for future development.

# 2. BACKGROUND

Historically, the development of synthetic data has been rooted in efforts to balance access to detailed data with the protection of confidentiality. For example, Rubin (1993) first proposed generating synthetic data as a disclosure-protection strategy, building on the multiple-imputation framework by replacing confidential values with values generated from statistical models (Rubin, 1978; Rubin, 1987). Subsequent research distinguished fully synthetic data in which all released values are generated, from partially synthetic data, in which selected sensitive values are replaced, and developed procedures for obtaining valid statistical inferences from multiple synthetic datasets (Raghunathan et al., 2003; Reiter, 2003; Reiter, 2005).

These early approaches to synthetic data generation relied primarily on specialized statistical methods and software, requiring researchers to explicitly select and configure the models used to generate synthetic observations (Nowok et al., 2016; Drechsler & Haensch, 2024). Because the synthetic process is explicitly specified, these approaches provide considerable control and make the assumptions underlying data generation relatively transparent. However, they require users to select appropriate models, specify relationships among variables, and configure the synthesis process, limiting the ease with which these methods can be adapted to a broad range of purposes.

The emergence of machine-learning methods has expanded the scope and complexity of data which can be generated synthetically. Generative adversarial networks (GANs), for example, use two neural networks that work against each other, where one network generates new samples and the other evaluates how closely those samples resemble the original data. (Goodfellow et al., 2014). Variational autoencoders (VAEs) likewise use neural networks, but differ in that their objective is to learn the underlying patterns in data and generate new observations based on those patterns (Kingma & Welling, 2014). Both approaches can capture complex relationships and distributions and are more readily adaptable across use cases than many specialized statistical methods. However, they typically require an existing dataset from which to learn and the resulting data-generating process is less transparent and more difficult for users to understand or control. They also introduce additional considerations related to model architecture, training, tuning, and evaluation (Xu et al., 2019; Little et al., 2025).

More recently, researchers have begun exploring large language models (LLMs), as another approach to synthetic data generation (Long et al., 2024; Miletic & Sariyar, 2024; Xu et al., 2024). LLMs provide an accessible and flexible interface, but their ability to generate statistically coherent tabular data cannot be assumed. Xu et al. (2024), for example, found that approaches that rely primarily on an LLM to generate synthetic observations may fail to preserve important relationships among variables, particularly in complex datasets. These findings underscore the importance of using LLMs within a structured process. For pedagogical purposes, such a process could combine the flexibility of LLMs with the transparency and control offered by explicitly specified statistical models. It should allow instructors to align the topic and statistical properties of a dataset with particular learning objectives while providing opportunities to inspect, validate, and refine the data-generating process. In the next section we describe how DagFlow-AI was designed to meet these needs.

# 3. SYNTHETIC DATA GENERATION USING DAGFLOW-AI

## 3.1 Technical Description

DagFlow-AI is a Python web application built with Streamlit and integrated with OpenCode. Its source code is maintained in a GitHub repository, and the application is configured to run through GitHub Codespaces, although it can be run anywhere that supports development container specifications. The application, source code, and instructions for accessing it are being withheld to preserve anonymity during the review process and will be included in the final version of the manuscript. DagFlow-AI integrates a coding agent harness known as OpenCode (Anomaly, 2026) with a specialized user interface for viewing agents' outputs. The user interacts with an agent orchestrator known as the "Synthesizer" which manages a "team" of agents to create realistic synthetic datasets designed around specific learning objectives and dataset requirements. The user works with the Synthesizer to develop the information needed to generate the dataset.

The six stages for this process are illustrated in Figure 1 and described in Table 1.

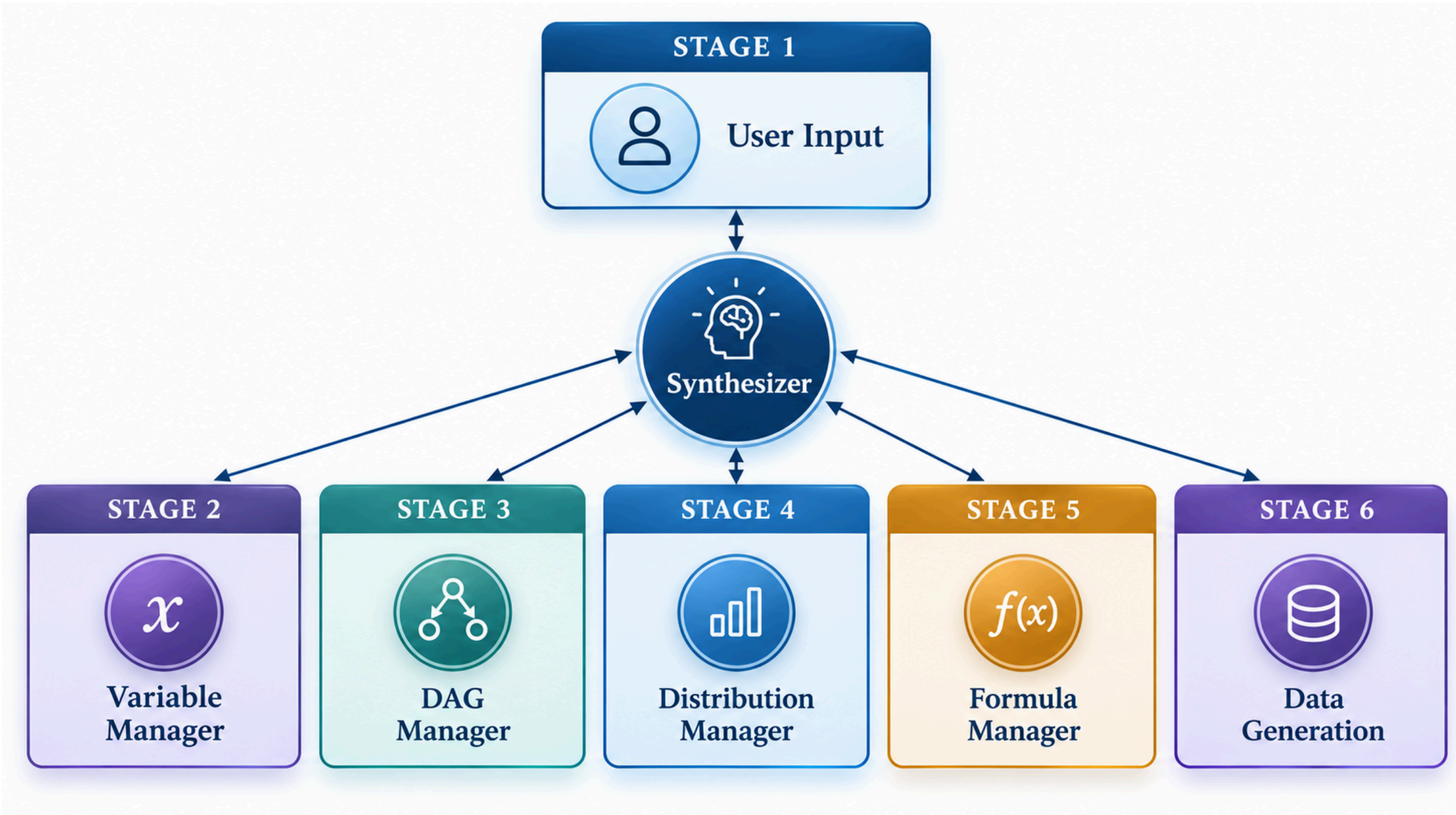

**Figure 1**: Illustration of synthetic data generation workflow for DagFlow-AI. The Synthesizer sits at the center as an orchestrator and delegates work to agents. Bidirectional arrows illustrate feedback loops between agents and Synthesizer as well as between Synthesizer and user.

**Table 1**: Description of the six stages of the DagFlow-AI synthetic data generation tool

| Name of Stage | Tasks Performed |
| --- | --- |
| 1: Dataset Description | The user provides the Synthesizer with the information needed to manage the dataset creation workflow. First, the user specifies a domain of interest or research question that provides context for the dataset. Second, they identify the learning objectives that the dataset should support. Finally, they provide any additional details or desired features. |
| 2: Variable Creation | A “Variable Manager” agent uses the information provided by the Synthesizer to create a list of relevant variables. It ensures that the variables support the desired learning objectives and meet other specified requirements. For each variable, it also provides metadata that can be used by other agents. |
| 3: DAG Creation | The “DAG Manager” uses the information provided by the Synthesizer and the variable list created by the Variable Manager to construct a directed acyclic graph (DAG). The graph specifies how variables influence one another and is intended to produce realistic relationships and associations in the final dataset. |
| 4: Distribution Selection | The “Distribution Manager” inspects the DAG and variable list and uses this information to select an appropriate distribution for each variable from a list of supported distributions. It also specifies feasible bounds for variables and selects realistic parameters for the distribution. This stage defines how each variable should be distributed once data generation is complete. |
| 5: Formula Generation | A “Formula Manager” defines a realistic linear or generalized linear model for each endogenous variable in the DAG. These models specify how the parent variables should influence each endogenous variable. The Formula Manager selects coefficients intended to represent plausible real-world relationships and to be appropriate for the type of model implied by the distribution of the response variable. |
| 6: Dataset Simulation | In the final stage, the Synthesizer runs a data-generation script that combines the information produced by the different agents and implements the resulting data generating model. The DAG determines the order in which variables are generated. Exogenous variables are sampled directly from their selected distribution, while endogenous variables are generated using the specified model formulas. If incompatible model specifications prevent successful simulation - for example, invalid distributional bounds or coefficients that produce numerically unstable values - the error is reported to the Synthesizer, which routes it to the appropriate agent for correction. This process continues until the dataset is successfully generated. |

Several suggestions from Sapkota (2025) regarding best practices for developing tools with agentic AI were followed. After each stage is completed or updated, the user can click on the corresponding tab in the user interface to inspect the output. This allows the user to see the choices made by the agent. The user can then provide feedback to the Synthesizer about any desired changes or updates, and the Synthesizer can route the request to the appropriate agent. If a request requires multiple pieces of information to be updated, the Synthesizer works to ensure that each change is made. It can also answer questions about the output produced at each stage or about the completed dataset.

In addition to being flexible, DagFlow-AI is designed to be reliable. This is achieved by separating tasks that should be completed by agents from those that should be completed programmatically. Agents are not used to generate the data directly or to write the code that generates the dataset. Rather they utilize their language understanding and reasoning skills, along with their broad knowledge of the world, to provide the pieces of information that collectively define a data generating model[1]. The Synthesizer regularly runs a validation script to ensure that agent outputs conform to the structure expected by the dataset-generation script. When errors occur, messages are passed back to the Synthesizer which can then determine how to resolve them. This approach reduces problems arising from agents producing unsupported output, generating erroneous code, or failing to follow instructions as expected.

Because DagFlow-AI is built using OpenCode and Streamlit in Python, it is not tied to a specific model provider. The Synthesizer can be configured to work with models from providers such as Anthropic or OpenAI, or through services such as OpenRouter or OpenCode Go. These options generally require an API key and may incur provider-specific usage charges. Alternatively, users can configure DagFlow-AI to work with locally hosted, open-weight models. When the application is run through GitHub Codespaces, users may also incur Codespaces charges, depending on their account plan and usage, although such charges should be rare.

[1] Agent models are trained on as much text as they can get as cheaply as possible, including scientific papers. A consequence of this is that while the abstracts for nearly all papers will be in their training sets, the full text will only be included for open access papers.

## 3.2 Pedagogical Alignment of DagFlow-AI

Although DagFlow-AI is primarily designed for instructors to create targeted, personalized learning experiences for students, its workflow and outputs can also be used to engage students in different forms of statistical thinking. We draw on two frameworks to describe these possibilities. The Problem, Plan, Data, Analysis, and Conclusion (PPDAC) framework characterizes the statistical investigative process (Wild & Pfannkuch, 1999). Although the first component is sometimes presented as Question rather than Problem (Steiner & MacKay, 2026), we use PPDAC here and treat the formulation of a research question as part of the Problem stage. We also draw on the recommendations and student learning outcomes from the Guidelines for Assessment and Instruction of Statistics Education (GAISE, 2015; ASA, 2025) summarized in Tables 2 and 3.

**Table 2**: Recommendations (R) from GAISE

| | |
|---|---|
| R1 | Teach statistics and data science as processes of gleaning insights from data to inform evidence-based decisions. |
| R2 | Emphasize effective written and oral communication of results from data, with attention to the scope and limitations of conclusions. |
| R3 | Focus on conceptual understanding rather than algebraic manipulation and formulas. |
| R4 | Integrate real data with a context and purpose throughout the course. Select data that are meaningful and engaging to the students. |
| R5 | Encourage multivariable thinking. |
| R6 | Incorporate software/apps to explore concepts and work with data. |
| R7 | Emphasize responsible and ethical conduct in the collection and use of data and in their analysis. |
| R8 | Employ evidence-based pedagogies that actively engage students in the learning process. |
| R9 | Use a variety of formative and summative assessments to improve teaching and learning. |
| R10 | Implement inclusive strategies that create learning environments where all students are valued. |

**Table 3**: Student Learning Outcomes (SLO) from GAISE

| | |
|---|---|
| SLO1 | Identify cases, variables, and types of variables in data sets, including multivariable data sets. |
| SLO2 | Explain how study design affects what conclusions can be drawn from the data. In particular, explain the importance of random sampling and random assignment. |
| SLO3 | Identify and interpret appropriate graphs and summary statistics in both univariate and multivariate settings. |
| SLO4 | Use statistical software or apps to create visualizations, produce summary statistics, and carry out the computational aspects of statistical analysis. |
| SLO5 | Articulate the role that random variation plays in statistical inference. In particular, articulate its role in generalizing from a sample to a population. |
| SLO6 | Identify the key methods to use in statistical inference and interpret results from those methods. |
| SLO7 | Recognize that sampling variation is one reason that conclusions drawn from statistical inference might be incorrect, and that replication and reproducibility are important to verify conclusions. |
| SLO8 | Use statistical models, such as linear regression, for prediction. |
| SLO9 | Recognize ethical issues and dilemmas in statistical practice. |
| SLO10 | Communicate, clearly and in context, results obtained from data. |

DagFlow-AI can first support thinking about how statistical investigations begin. In Stages 1 and 2, users identify a topic, formulate a problem or research question, specify the intended learning objectives, and determine which variables are needed. These activities relate most directly to the Problem and Plan components of PPDAC. Instructors can involve students in deciding what information would be needed to address a question, identifying the cases and variables, and considering whether the proposed variables are relevant to the problem (SLO1). Topics can also be selected to reflect students' interests or fields of study. Although synthetic data do not replace the real data emphasized by GAISE, they can support its broader emphasis on purposeful, meaningful, and engaging settings (R4).

Stages 3 through 5 provide additional opportunities to reason about how data arise. The directed acyclic graph makes proposed relationships among variables explicit, while the selected distributions and model formulas specify how those relationships will be represented in the generated data. Students can inspect, critique, or revise these components rather than treating a dataset as a collection of observations whose origins are unknown. For example, they can consider whether important variables or relationships are missing, whether a proposed association is plausible, and whether potential confounders have been represented appropriately. These activities support conceptual understanding (R3), multivariable thinking (R5), and reasoning about how the structure of a study affects the conclusions that can be drawn (SLO2).

The distributions and formulas chosen by the agents can also be used to develop students' understanding of statistical models. Students can determine whether a distribution is appropriate for a particular type of variable, anticipate the shape and variability of the resulting observations, and identify suitable summaries or visualizations (SLO1, SLO3). Model formulas can be used to explore how predictors influence a response, how coefficients encode the direction and strength of relationships, and how several variables can operate together in a regression model (SLO8). Asking students to explain or critique these choices emphasizes the concepts represented by a model rather than only its algebraic form (R3).

Once observations are generated in Stage 6, the dataset can support the Data, Analysis, and Conclusion components of PPDAC. Students can use statistical software to create visualizations and numerical summaries, fit models, conduct inferential procedures, and interpret the results (R6, SLO3, SLO4, SLO6, SLO8). They can then communicate what the data suggest while attending to the scope and limitations of their conclusions (R1, R2, SLO10). Because the underlying data-generating model is available, instructors can also ask students to compare the relationships they infer from a sample with the relationships used to generate it.

The ability to generate repeated datasets creates further opportunities to study variation. Students can generate multiple samples from the same data-generating model and examine how sample statistics, parameter estimates, model results, and conclusions differ across samples. They can also modify features of the model and investigate how those changes affect the resulting data. These activities make sampling variation, replication, and reproducibility more concrete (SLO5, SLO7).

DagFlow-AI can also support discussion of responsible statistical practice. Instructors can design scenarios involving measurement choices, omitted variables, confounding, or other ethical concerns and ask students to identify their consequences (R7, SLO9). More generally, inviting students to question and revise the agents' decisions positions AI-generated output as something to be evaluated rather than accepted uncritically. In this way, the tool can support active engagement throughout the investigative process (R8), whether students work with only the final dataset or participate in constructing and evaluating the data-generating model itself.

# 4. CASE STUDY

This section presents a case study demonstrating how DagFlow-AI can be used to develop, inspect, and refine a synthetic dataset tailored to a student's interests and the learning objectives of an introductory statistics course. Figure 2 shows the DagFlow-AI interface, with the OpenCode chat terminal on the left and tabs for viewing the outputs from each stage on the right.

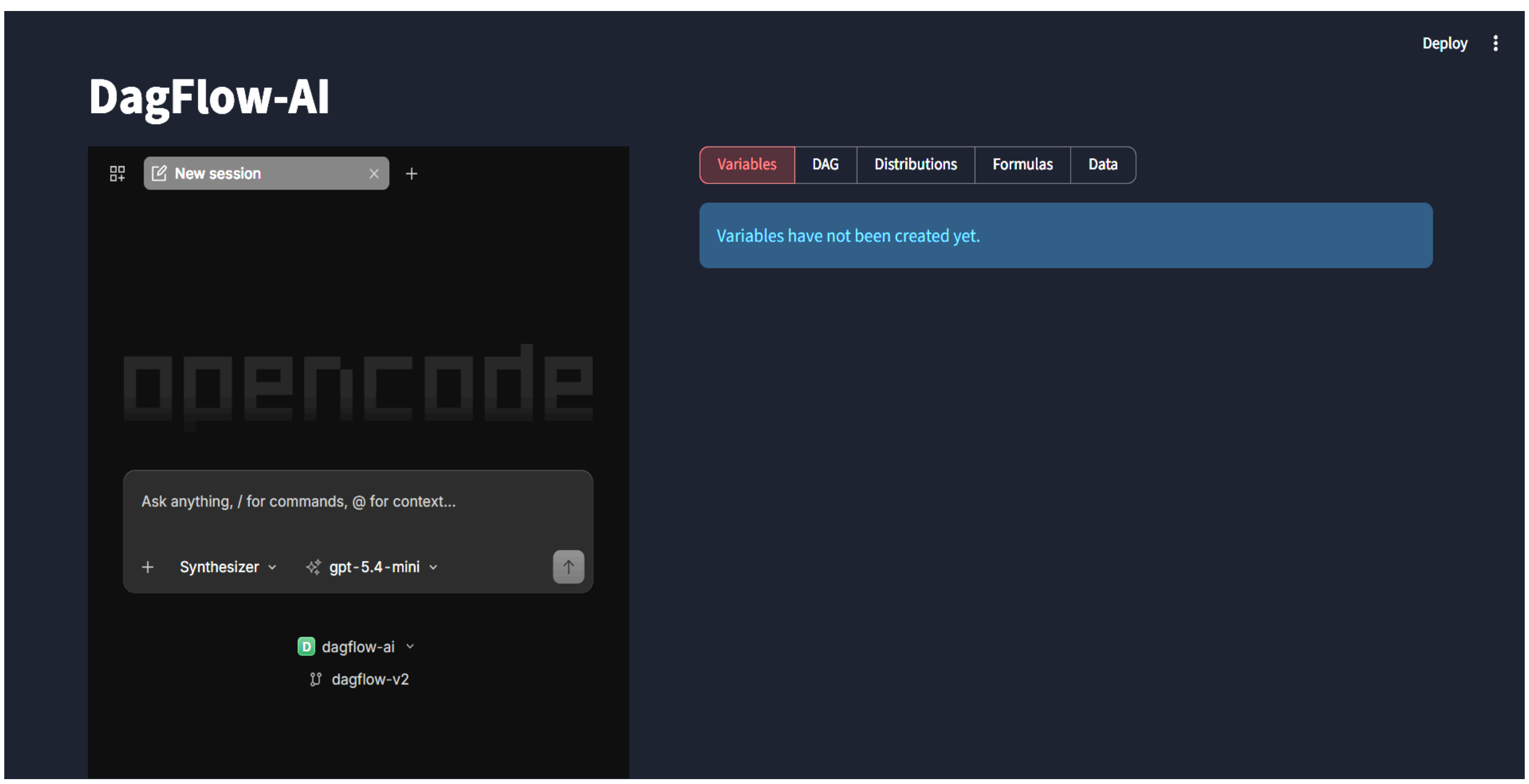

**Figure 2**: Web application for DagFlow-AI. OpenCode chat terminal can be found on the left and viewers are found by clicking on tabs on the right.

To begin the synthetic dataset generation, the user provides three pieces of information:

1. A problem or question that establishes a domain of study or intended application.
2. The learning objectives the dataset should support.
3. Any additional details or requirements for the dataset.

Consider a student who has expressed the following interests: "I am very interested in the environment, climate change, and what we can be doing to minimize the impact of climate change on the environment and our health". The student is enrolled in a calculus-based, introductory statistics course covering exploratory data analysis, statistical distributions, confidence intervals, and simple linear regression. The prompt provided to the Synthesizer to begin the data-generation process is shown in Figure 3.

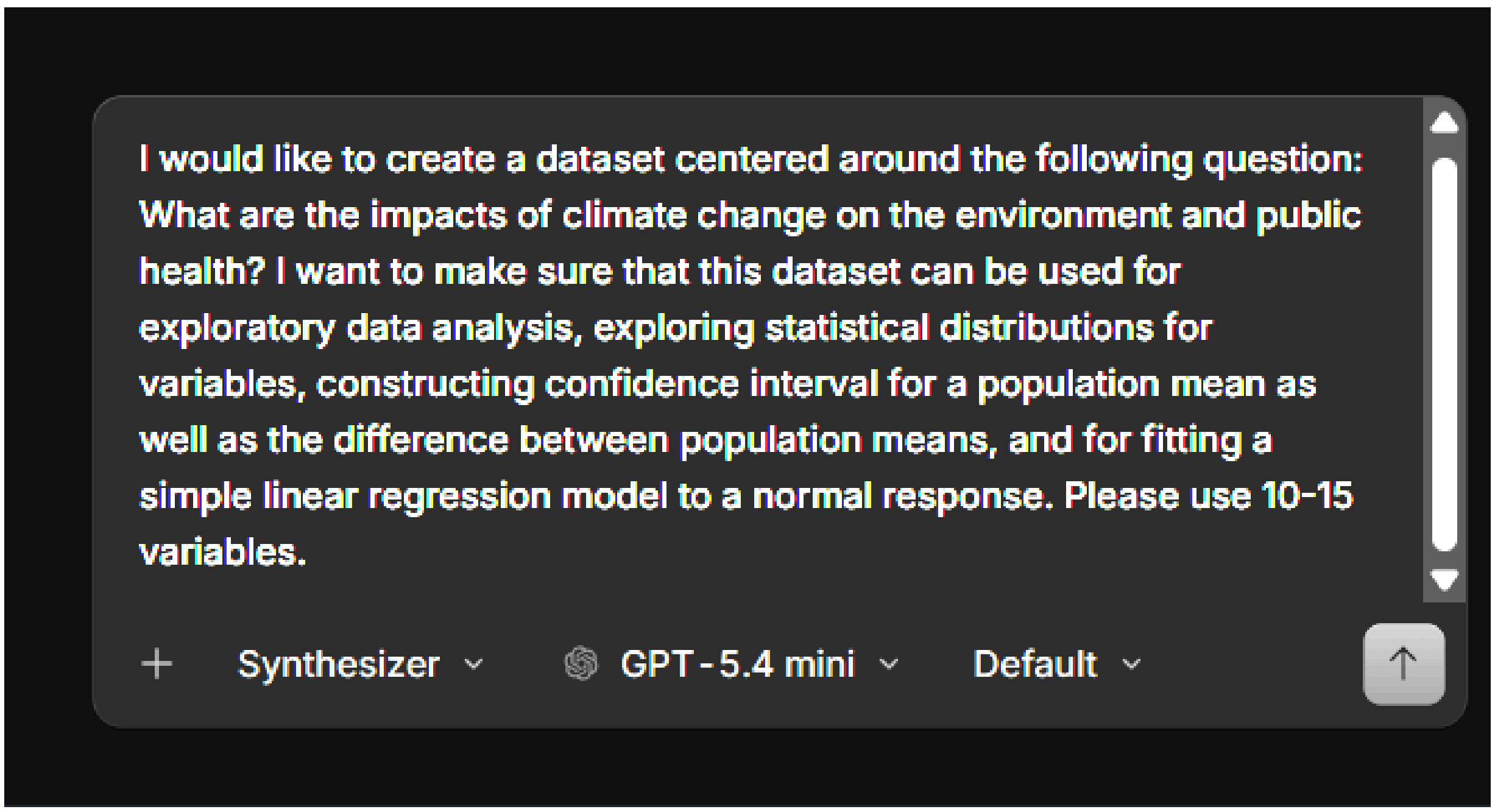

**Figure 3**: Example of prompt used to begin the synthetic data generation process with the DagFlow-AI Synthesizer

The Variable Manager selects a wide variety of variables that fit this setting, several of which are shown in Figure 4. The list includes discrete and continuous quantitative variables with different distributional shapes, as well as categorical variables. Quantitative variables can be used to construct confidence intervals for population means, while binary categorical variables can be used to construct confidence intervals for population proportions. A binary categorical variable can also serve as a grouping variable when constructing a confidence interval for a difference in population means.

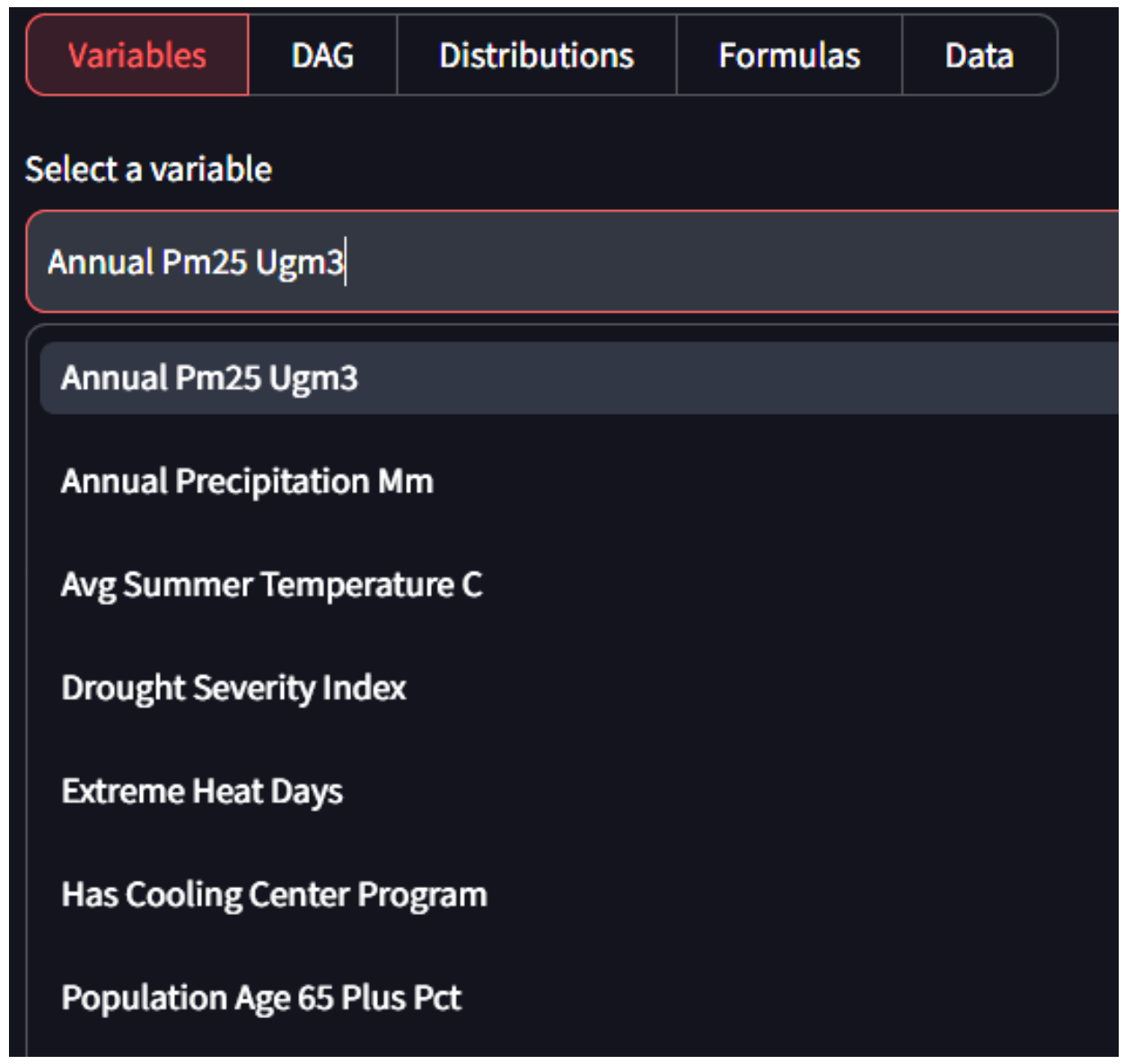

**Figure 4**: A selection of variables created for the climate, environment, and health example

Several variables could be used as the response variables in a regression model. The Variable Manager appears to have selected "Respiratory Hospitalization Rate Per 100K" for this role. As shown in the variable viewer in Figure 5, this variable is specified as right-skewed. The instructor could ask for this specification to be changed or retain it as an opportunity for students to assess regression assumptions and consider whether a simple linear regression model is appropriate. Alternatively, the instructor could adjust the distributional parameters or select a variable with less skew, such as "Drought Severity Index," as the response.

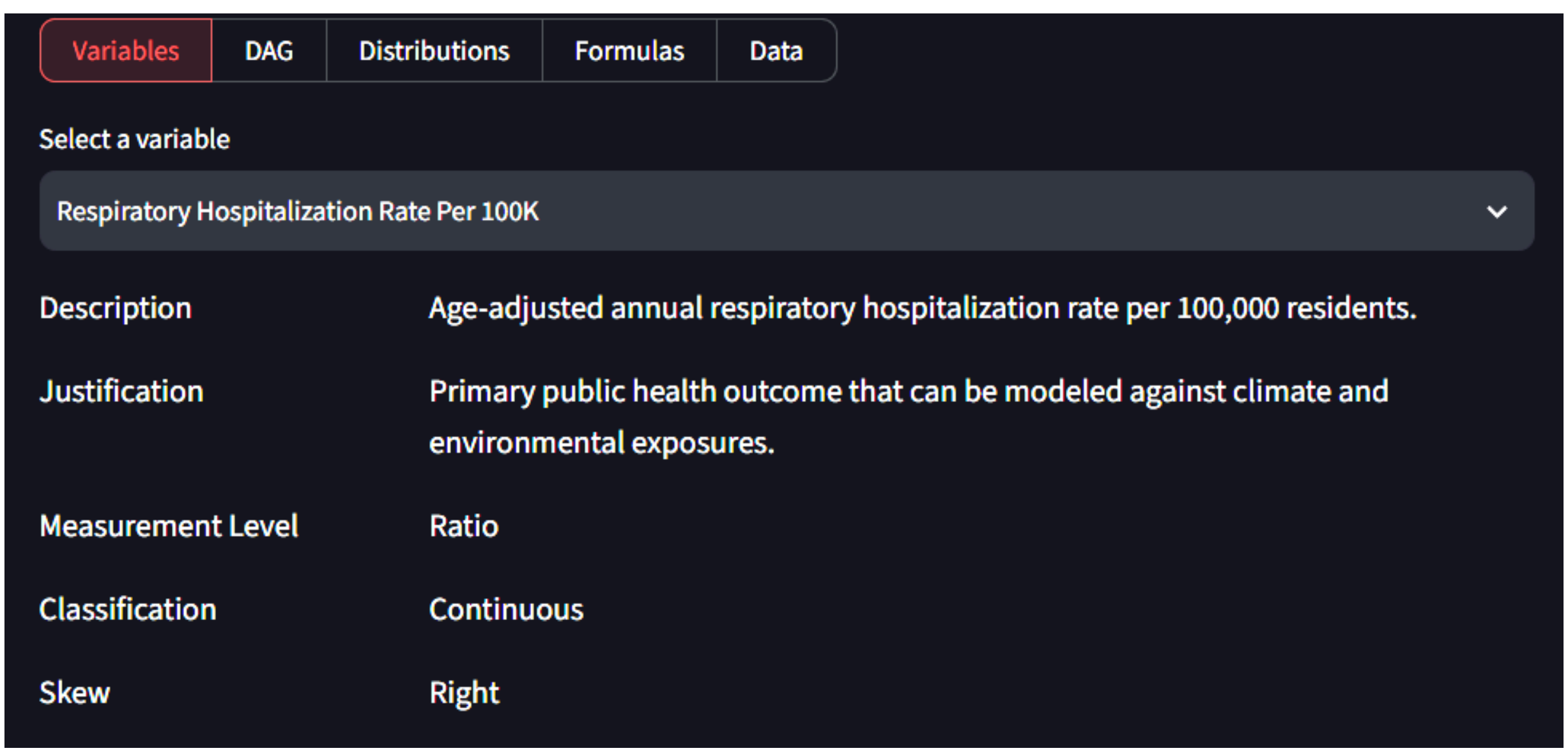

**Figure 5**: Variable viewer for "Respiratory Hospitalization Rate Per 100K"

Using this variable list, the DAG manager constructs a directed acyclic graph representing the proposed relationships among the variables. The resulting DAG is shown in Figure 6. Six variables are identified as direct influences on “Respiratory Hospitalization Rate Per 100K” and these relationships appear to be grounded in the relevant literature (Trachtenberg et al., 2014; Vaidyanathan et al., 2019; Vega et al, 2025). The instructor can use the DAG as generated, request modifications before generating the dataset, or invite students to reason about and critique the agent’s choices. For example, students could discuss how “Urbanicity” and “Region” might serve as confounders or consider the ethical and societal implications of the greater vulnerability of certain populations.

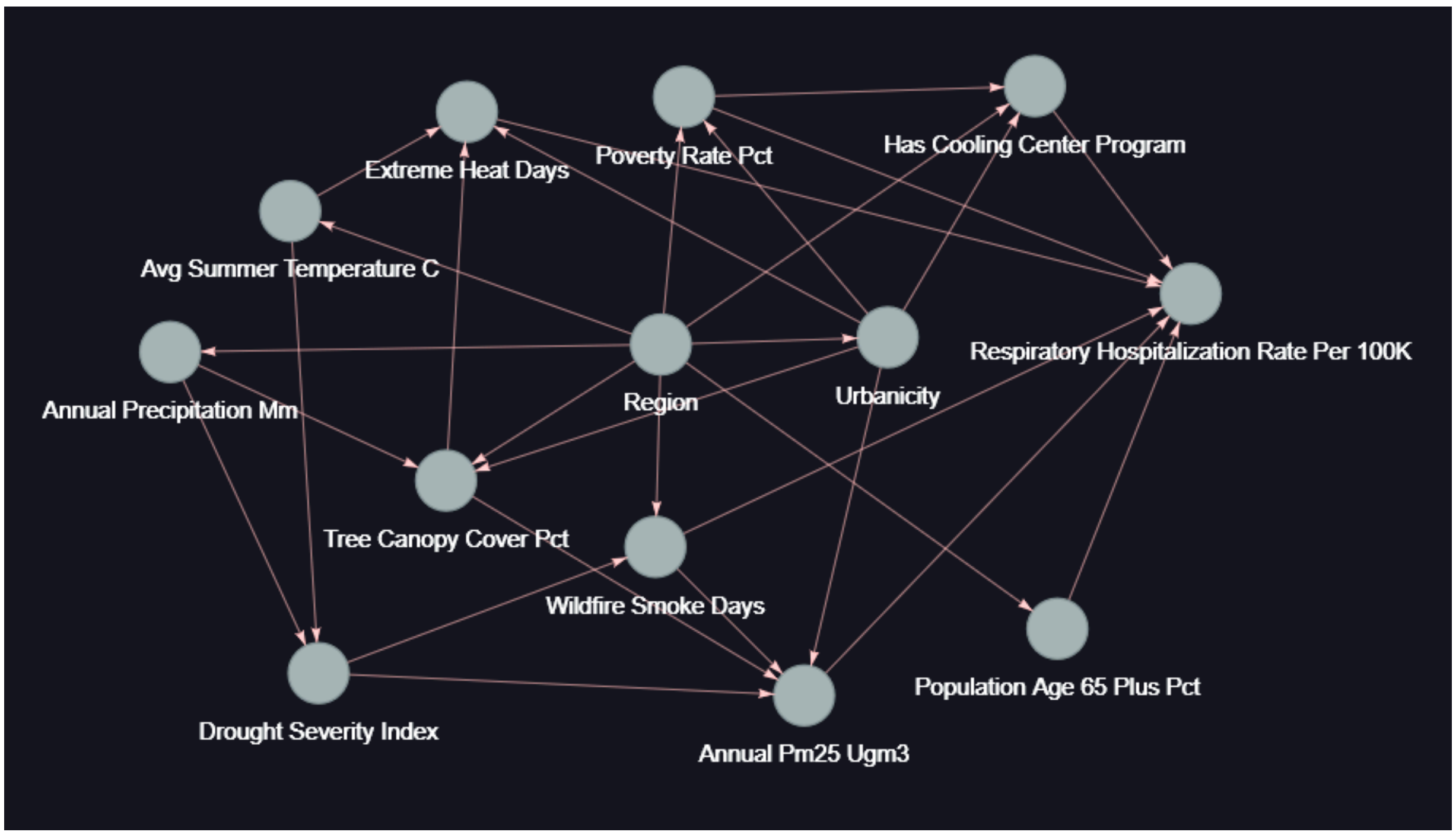

**Figure 6**: Directed acyclic graph created by the DAG manager for the climate, environment, and health example

In the next stage, the Distribution Manager selects a distribution for each variable. The instructor can inspect these choices and request adjustments as needed. The resulting specifications include categorical variables and both discrete and continuous quantitative variables, making the dataset suitable for exploratory data analysis and exercises in selecting appropriate probability distributions. Students in more advanced courses could evaluate the agent’s choices and modify distributional parameters to improve realism. Figure 7 shows the selected distribution and parameters for “Respiratory Hospitalization Rate Per 100K” after adjustments were made to enable successful data generation.

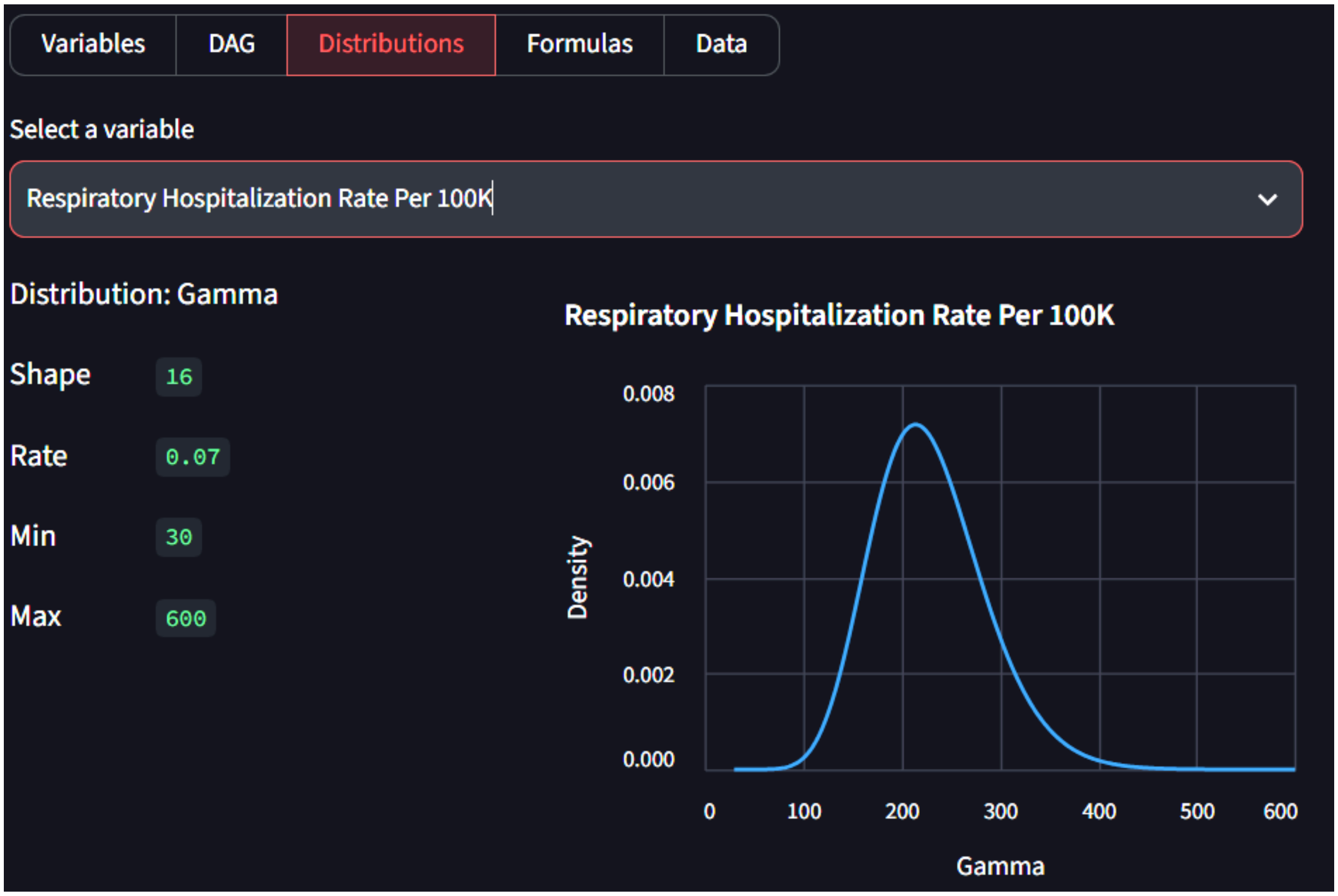

**Figure 7**: Distribution selected for "Respiratory Hospitalization Rate Per 100K" after adjustments were made to enable successful data generation.

In the penultimate stage, the Formula Manager proposes a formula for each endogenous variable in the DAG. These formulas specify the direction and magnitude of the relationships between each endogenous variable and its parents [2]. The final formula for "Respiratory Hospitalization Rate Per 100K" is shown in Figure 8. With the exception of the presence of a cooling center program, each factor in the formula is associated with an increase in respiratory hospitalization rates. The directions and relative magnitude of these relationships are largely consistent with findings in the relevant literature.

[2] The initial formulas shown in the box only represent the intended direction and relative magnitude of effects. Actual coefficient values will likely need to be adjusted during the dataset generation process to ensure sampled values are consistent with specified target distributions. Agents can assist the user with this process.

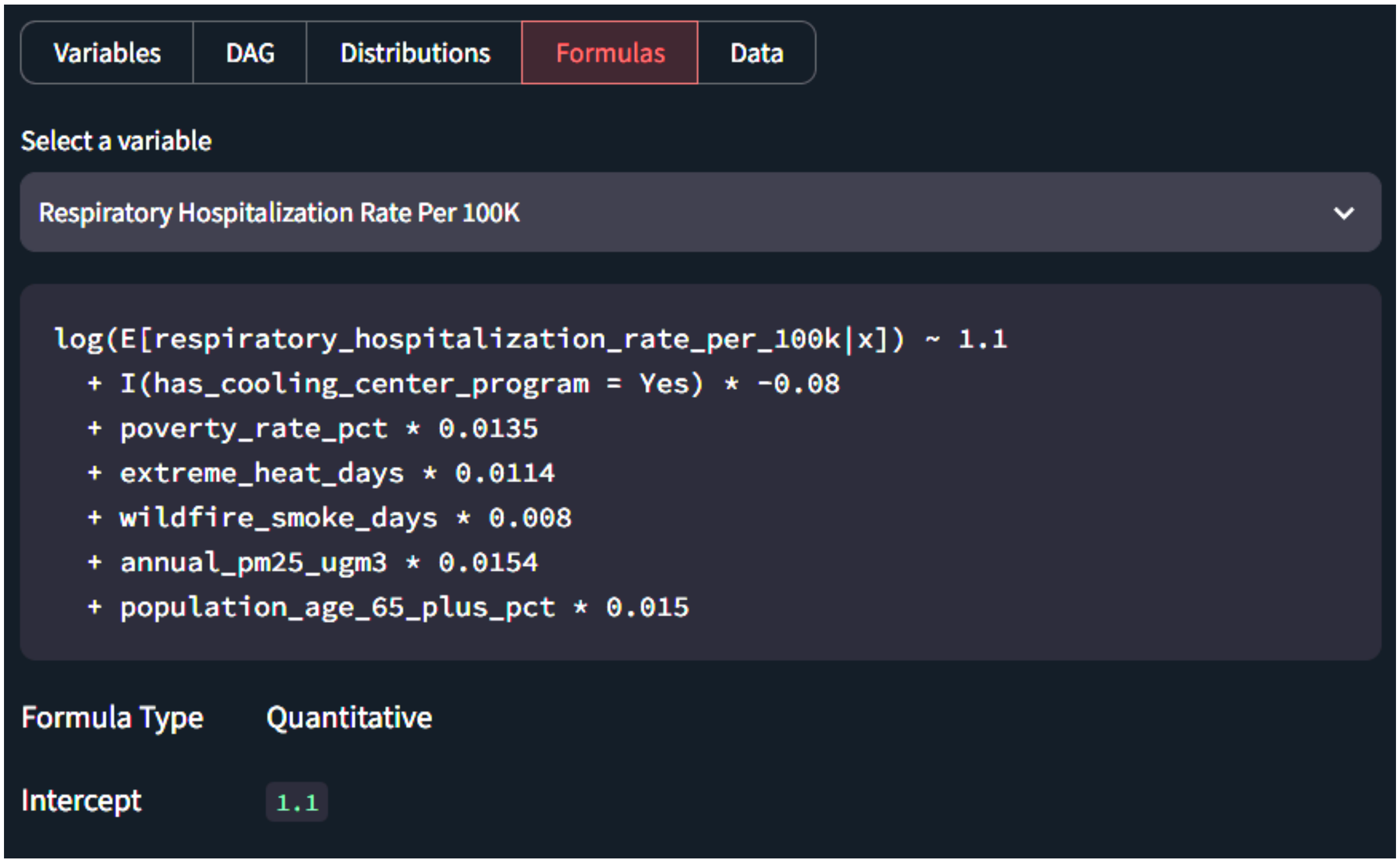

**Figure 8**: The final formula selected for "Respiratory Hospitalization Rate Per 100K" shown in the formula viewer. This formula was obtained after several iterations and adjustments to ensure stable dataset generation.

The instructor can work with agents to adjust formulas until the generated dataset exhibits the relationships intended for students to investigate. The formulas themselves also provide learning opportunities. Students in introductory courses can focus on the signs and relative magnitudes of the coefficients and reason about what the formula implies about relationships among variables. More advanced students can interpret the coefficients while accounting for the response distribution and link function. For example, a coefficient of 0.0114 for extreme heat days under a log link implies that each additional extreme heat day multiplies the expected respiratory hospitalization rate by $e^{0.0114} = 1.011$, corresponding to an increase of 1.1%, adjusted for the other variables.

Once any necessary adjustments to the distributions or formulas have been made, the synthetic dataset can be generated. The instructor can then use the visualization tools within the application to examine whether the resulting dataset behaves as expected. Figure 9 shows a moderate positive relationship between annual PM2.5 concentration and the respiratory hospitalization rate, making this variable a potential candidate for a simple linear regression activity. Figure 10 compares respiratory hospitalization rates for observations with and without cooling-center programs. The apparent difference between the groups could support an activity involving a confidence interval for the difference in population means.

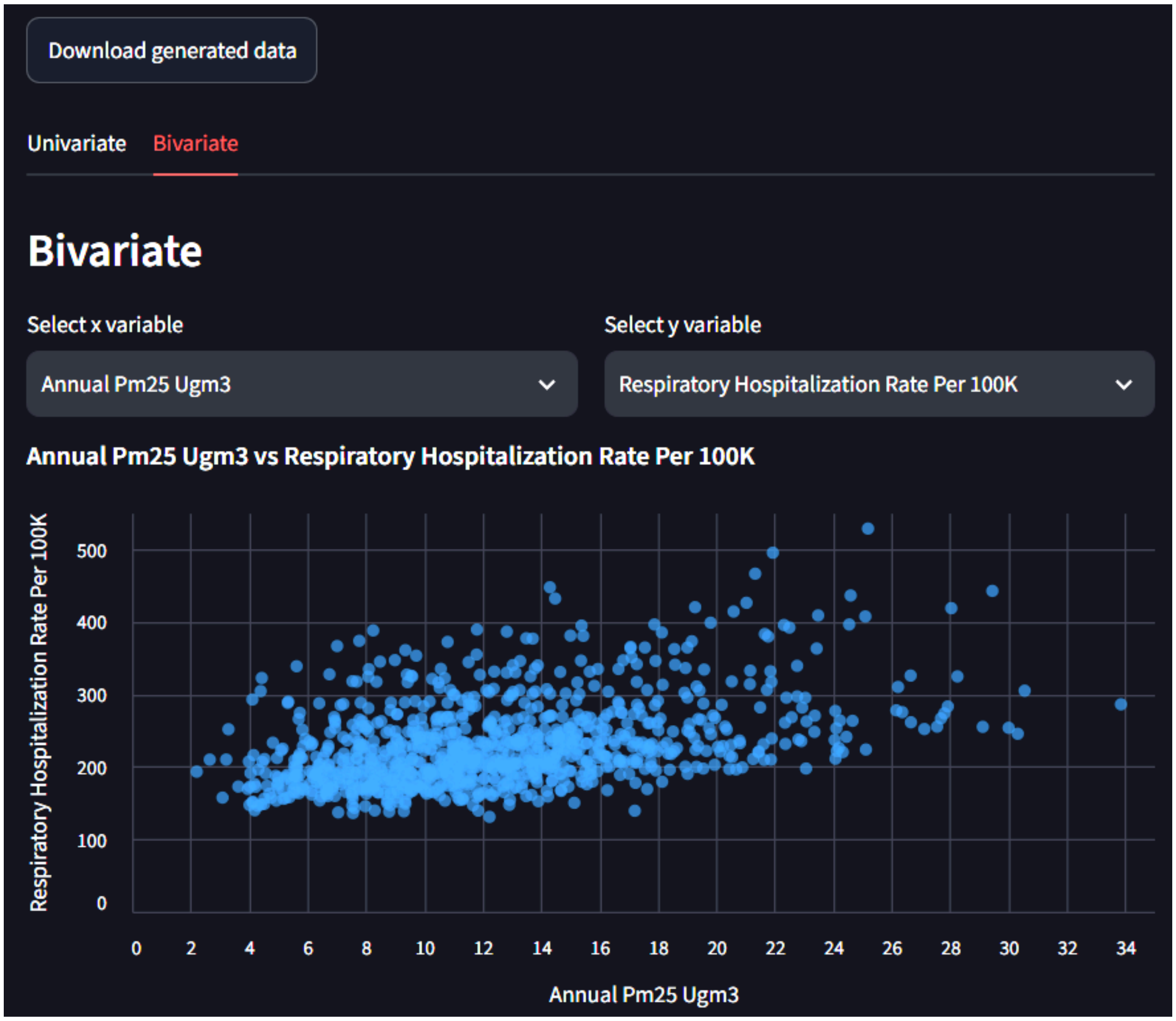

**Figure 9**: Bivariate visualization of "Annual PM25 UGM3" and "Respiratory Hospitalizations Rate Per 100K" produced from the synthetic dataset.

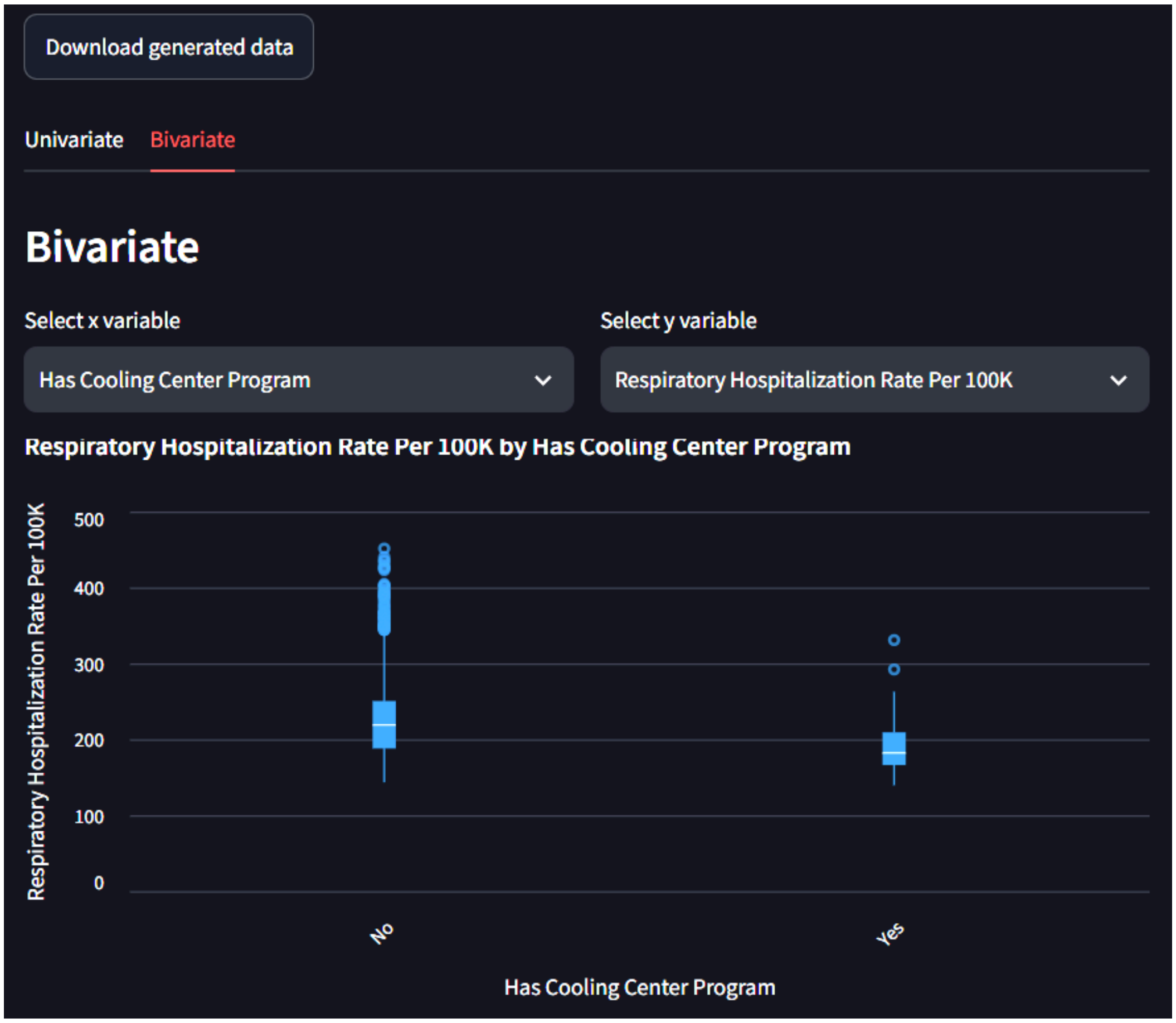

**Figure 10**: Bivariate visualization of "Has Cooling Center Program" versus "Respiratory Hospitalization Rate Per 100K" produced from the synthetic dataset.

# 5. DISCUSSION

This paper demonstrates how agentic AI can be used to create personalized synthetic datasets for statistics education. Existing approaches often require instructors to begin with a suitable real dataset, construct a data-generating process themselves, or use a tool designed for a particular statistical setting. DagFlow-AI addresses these constraints by using LLM-based agents to construct the components of an explicit, inspectable data-generating mechanism aligned with a chosen topic and learning objective. Programmatic procedures then validate the structure of those components and generate the observations.

DagFlow-AI is intended to complement rather than replace real datasets in the classroom. Because synthetic data do not arise from actual individuals, objects, or events, they cannot support substantive conclusions about the world or provide experience with all aspects of data collection, measurement, and study design. They can, however, allow instructors to control the context and statistical properties of a dataset, creating new opportunities for students to explore statistical ideas in varied and meaningful settings.

The tool can support at least two broad types of learning experiences. First, students can be given a generated dataset and asked to select and apply appropriate statistical methods, identify patterns and relationships embedded in the data, and communicate their findings. Second, students can inspect, critique, or modify the data-generating mechanism itself. This use encourages students to view statistical models as representations of real-world phenomena and to reason about their assumptions, strengths, and limitations.

The present work demonstrates the design and feasibility of this approach through a single case study. Building on this foundation, future technical development could incorporate relevant external information, support additional distributions and model classes, and expand the procedures used to evaluate the resulting data-generating mechanisms and datasets. We also plan to use DagFlow-AI in our own classrooms, experiment with different instructional uses, and evaluate both student learning and students' reactions to working with personalized synthetic data. This research could compare experiences in which students analyze a completed dataset, participate in constructing its data-generating mechanism, or work with both synthetic and real data. The findings will help clarify when personalized synthetic datasets are most pedagogically useful and how they can best complement students' experiences with real data.

# 6. ACKNOWLEDGEMENTS AND FUNDING

The authors thank Niki Burggraf (UW eScience Institute) for their guidance on workflow development, technical features, and improving our project's computing environment. This work was supported by the SEED-AI program via the AI@UW Initiative, funded by the Charles and Lisa Simonyi Launch Fund for Artificial Intelligence at the University of Washington.